\documentclass[useAMS,usenatbib]{mn2e}
\usepackage{epsfig}   
\input{psfig}
\def\<#1>{\langle\hbox{#1}\rangle}
\def\etal{{et~al.}}

\def\eg{{e.g.}}
\def\ie{{i.e.}}
\def\simgt{\hbox{\rlap{\raise 0.425ex\hbox{$>$}}\lower 0.65ex\hbox{$\sim$}}}
\def\simlt{\hbox{\rlap{\raise 0.425ex\hbox{$<$}}\lower 0.65ex\hbox{$\sim$}}}
 
\def\deriv(#1/#2){\mathchoice{d#1\over d#2}
    {d#1/d#2} {d#1/d#2} {d#1/d#2}}
\def\pderiv(#1/#2){\mathchoice{\partial#1\over\partial#2}
    {\partial#1/\partial#2} {\partial#1/\partial#2} {\partial#1/\partial#2}}
\def\pderivp(#1/#2){\left(\pderiv(#1/#2)\right)}

\def\tderiv(#1/#2){\mathchoice{d#1\over d#2}{d#1/d#2}
                   {d#1/d#2}{d#1/d#2}}
\def\tderivp(#1/#2){\left(\tderiv(#1/#2)\right)}
\title[The Pisces-Cetus Supercluster]{The Pisces-Cetus Supercluster: 
a remarkable filament of 
galaxies in the 2dF Galaxy Redshift Survey and Sloan Digital Sky surveys}
\author[Porter and Raychaudhury]{Scott C. Porter and
        Somak Raychaudhury\thanks{E-mail: scp@star.sr.bham.ac.uk;
              somak@star.sr.bham.ac.uk}\\
School of Physics and Astronomy, University of Birmingham, 
Birmingham B15~2TT, UK}

\begin{document}

\date{Received July 2005}

\pagerange{\pageref{firstpage}--\pageref{lastpage}} \pubyear{2005}

\maketitle

\label{firstpage}

\begin{abstract}
The Pisces-Cetus supercluster (redshift $z\!\sim\! 0.06$) 
 is one of the richest nearby
($z\!<\!0.1$) superclusters of galaxies, and emerges as a remarkable
filament of galaxies at the edges of the two-degree-field galaxy
redshift survey and the ongoing Sloan Digital Sky Survey.  We explore
the extent of the supercluster on the sky and in redshift space, and
map the distribution of its constituent clusters and groups.  We find
evidence of enhanced star formation in galaxies belonging to groups in
the supercluster compared to those in the field. This effect appears
to be higher among the poorer groups than in the richer ones.  In
contrast, star formation is suppressed in the galaxies in rich clusters,
which is consistent with previous studies.  We identify two major
filaments in this supercluster, consisting of 11 and 5 Abell clusters,
including Abell~133 and Abell~85 respectively,
and estimate the virial masses of the clusters from their velocity
dispersions and optical surface brightness profiles. The
lower limit to the masses of these filaments, amounting to the
total virial mass of the constituent clusters, turns out to be $5.2
\times 10^{15}\,h_{70}^{-1}\,M_{\odot}$ and $6.0
\times 10^{15}\,h_{70}^{-1}\,M_{\odot}$ 
over volumes of almost $6750\,h_{70}^{-3}$~Mpc$^3$ and
$19600\,h_{70}^{-3}$~Mpc$^3$ respectively.  This corresponds to mass
overdensities of $\Omega/\Omega_{crit}\equiv\delta_M>4.7$ and $\delta_M>1.3$ for the two
filaments making up the supercluster.
\end{abstract}

\begin{keywords}
Galaxies: clusters: general;  Galaxies: evolution; Cosmology: large-scale 
structure of the Universe; X-rays: galaxies: clusters.
\end{keywords}

%=section1===================================

\section{Introduction}

Observations and simulations of the large-scale structure of the
Universe have revealed the presence of a network of filaments and
voids in which most galaxies seem to be found
\citep{zucca93,einasto94,virgo98}. This implies that that the Universe
is not homogeneous on scales below $\sim 100\,h^{-1}_{70}$ Mpc
\citep{bbs04,sss04}, which is the scale of the largest common
structures of galaxies, though the discovery of structures far larger
than these have been claimed 
\citep[e.g.,][]{bagchi02,brand03,miller04}.  
Large-scale structures
that are much larger than the virial radii of individual rich clusters
($\simgt 10\,h^{-1}_{70}$ Mpc) are often called superclusters of
galaxies, even though this term has been used to denote various
entities in the literature. We adopt this terminology in this paper.

Superclusters are the largest gravitationally bound systems in the
Universe, though it is not clear to what scales this property
extends. They are an essential tool to study the largest-scale density
perturbations that have given rise to structure in the
Universe \citep[e.g.,][]{neta00}. These are particularly
interesting since they have always evolved in the linear regime and
can be analytically tractable. Superclusters can be useful in
quantifying the the high-end  mass function of collapsing systems 
and ratio of mass to light on the largest scales, thus being useful in 
discriminating between dark matter and structure formation models
\citep[\eg,][]{kolo02,neta88}.

\begin{figure*}
%\centerline{\psfig{figure=groupnclus.ps2,width=4.0truein,angle=-90}}
%\centerline{\psfig{figure=groupnclus_bw.ps,width=0.8\hsize,angle=0}}
\centerline{
\psfig{figure=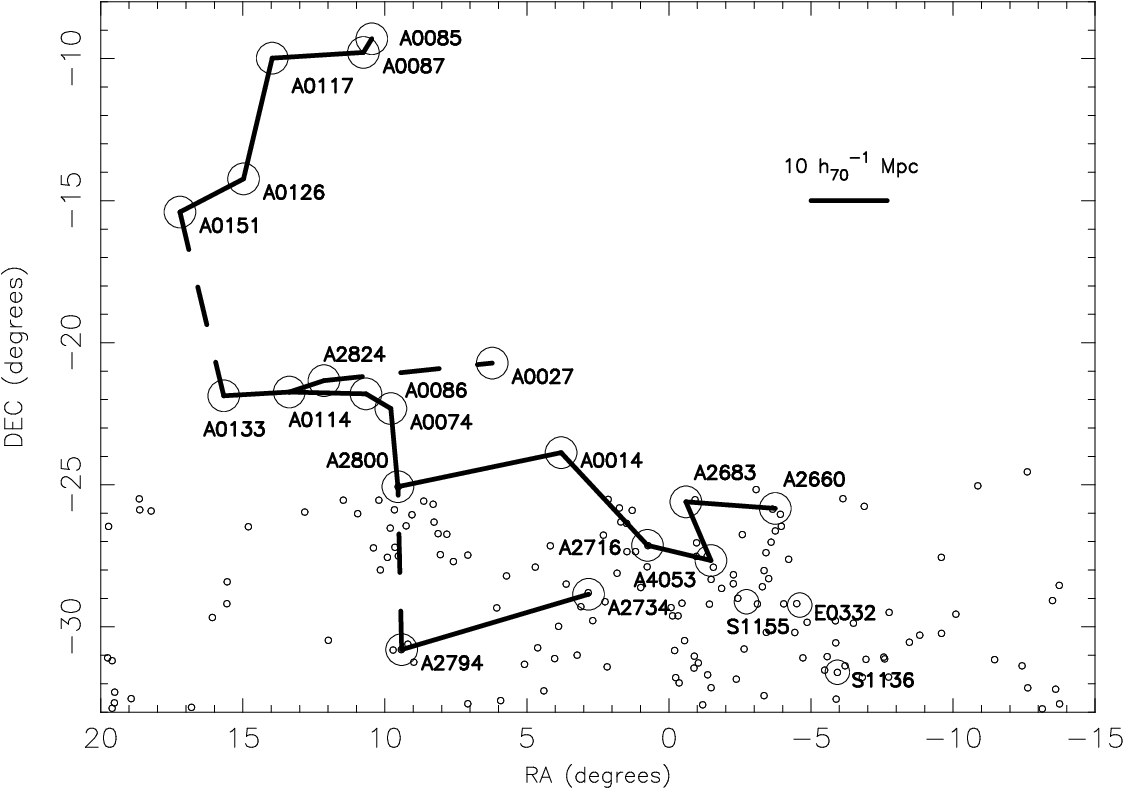,width=0.48\hsize,angle=0}
\psfig{figure=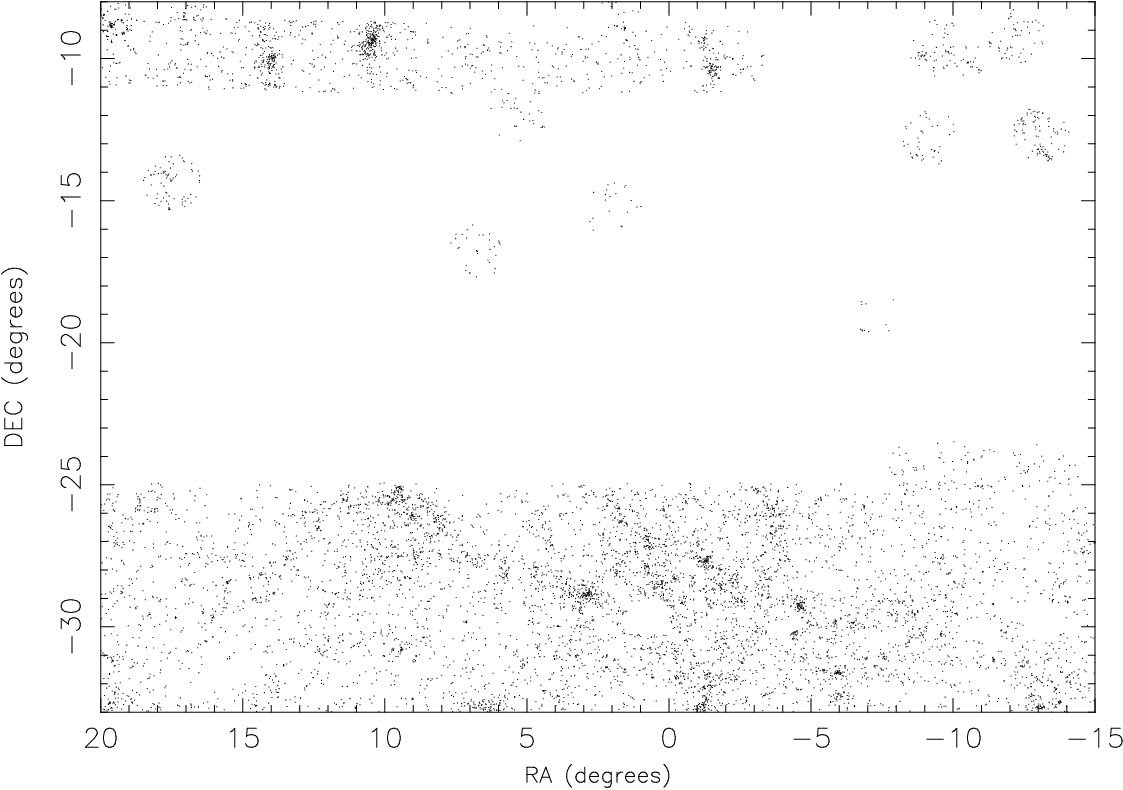,width=0.48\hsize,angle=0}
}
\medskip
 \caption{ (a, Left) Clusters of galaxies belonging to Pisces-Cetus
 supercluster. The minimal spanning tree used to identify the
 supercluster \citep{somak05} is shown connecting the clusters. Edges
 in the tree with length $20\,h^{-1}\!<\! L \!\le\! 25\,h^{-1}$ Mpc
 are shown as dashed lines. The groups of galaxies (2PIGG) identified
 from a friends-of-friends analysis of the 2dFGRS \citep{eke2004} are
 shown as little circles.  (b, Right) All galaxies with measured
 redshifts from the 2dFGRS and SDSS surveys in the region, between
 redshifts of 13920 and 22590 km/s. The SDSS reaches down to $\delta=-10^\circ$
in this RA range, and covers the clusters A85, A87 and A117. The SDSS 
reaches up to $\delta=-25^\circ$, and thus a large part of the supercluster
falls in the gap between the two surveys, except for the occasional isolated 
two-degree fields of the 2dFGRS.
\label{fig:tree}
}
\end{figure*}

Superclusters are expected to be predominantly filamentary structures.
Numerical simulations of the evolution of
large-scale structure \citep[\eg,][]{bond1996,colberg2005}, 
as well as
the analyses of large-volume surveys like the 2dfGRS 
\citep[\eg,][]{pimbblet2004} or studies of individual systems of
superclusters \citep[\eg,][]{bagchi02,ebe04}
show the pre-eminence of filamentary structures.
Nevertheless, supercluster catalogues are usually constructed
\citep[\eg,][]{einasto94,zucca93} using percolation or friends-of-friends
algorithms which do not take this geometrical feature into
consideration.

Recently, \citet{somak05} have compiled a catalogue of
supercluster-size structures from a redshift survey of Abell clusters
($z\!<\!0.1$) using the minimal spanning tree statistic, which
contains all the information used in a percolation analysis
\citep{bs96}, and is particularly sensitive to filaments.
 In this catalogue,
only one of the long filaments (containing 8 or more Abell clusters) is
in the 2dF galaxy redshift survey \citep{2dF-final} [2dFGRS] or Sloan Digital
Sky Survey \citep{sdss} [SDSS]
regions. In fact, it straddles both of them.

More than a decade ago, in a series of papers,
\citet{tully86,tully87,tully88} pointed out the Pisces-Cetus
supercluster to be one of the five richest systems of clusters of
galaxies in the $z\!<\! 0.1$ Universe, and speculated that it could be
as large as $300\, h_{70}^{-1}$ Mpc across, and might include the
Local Group. Subsequent work \citep[\eg,][]{zucca93,einasto94} has
pointed out interesting features of significantly smaller
substructures of this remarkable web of galaxies, groups and clusters,
and at least one unusually large void \citep{burns88}.  

In the supercluster catalogue of \citet{somak05}, the Pisces-Cetus
supercluster occupies a region of about 700 sq. deg. of the
(equatorial) southern sky. While using the Minimal Spanning Tree (MST) 
in constructing this
catalogue, if a maximum edge-length of $20\,h_{100}^{-1}$ Mpc is used, the
system breaks up into two filaments (A, B) of 11 and 5 clusters
respectively, at mean redshifts of 0.0625 and 0.0545 respectively. If,
however, adjacent clusters on the tree are allowed to be separated by
a distance of up to $25\,h_{100}^{-1}$ Mpc, the supercluster becomes a long
filament of 19 clusters at a mean redshift of 0.0591, making the
supercluster the third richest in the catalogue (following the Shapley
and Horologium-Reticulum superclusters). This is illustrated in
Fig.~\ref{fig:tree}a, where the clusters excluded by the former
consideration are shown to be connected to the main tree by dashed
lines.

In this paper, we seek to characterise the extent of the supercluster
in \S2 from the 2dFGRS and SDSS surveys, and from the SuperCosmos
Sky Survey
\citep{scos}. In  \S3 we calculate mean velocities and velocity
dispersions for the constituent clusters, while in \S4 we look at
evidence of enhanced star formation in galaxies belonging to groups
along the filaments of the supercluster, as an indicator of 
galactic interaction in superclusters. In \S5 we calculate
virial masses of the constituent clusters based on 
their optical luminosity profiles and velocity dispersion,
to estimate a lower limit to the mass overdensity in the
supercluster. We summarise our major 
conclusions in \S6. Detailed notes and X-ray images of the constituent
clusters can be found in the Appendix.
Throughout this paper we use $H_0\!=\!70$~km~s$^{-1}$ Mpc$^{-1}$.

%=section2=======================================================

\section[]{Clusters of galaxies belonging to the Pisces-Cetus Supercluster}

The clusters comprising the Pisces-Cetus supercluster from the MST
analysis of \citet{somak05} form the core of our sample, and are
listed in Table~\ref{tab-list}.  Unlike MST analyses of galaxy
catalogues, where completeness is an important consideration, the
supercluster list obtained in this catalogue is considered reasonably
complete since it arises from an analysis of rich Abell clusters in
the volume closer than $z<0.1$ (Galactic latitude $|b|>20^\circ$).

To the member list of the Pisces-Cetus Supercluster from the above
list, We added a few clusters, in the same volume as the filament,
from the Supplementary Abell cluster catalogue \citep{Abell1989} and
the Edinburgh-Durham Southern Galaxy Cluster Catalogue
\citep{Lumsden1997}. These extra clusters were taken 
as potential members of the supercluster  if 
their redshift as derived in \S3 was found to be
consistent with the other members.

\begin{figure}
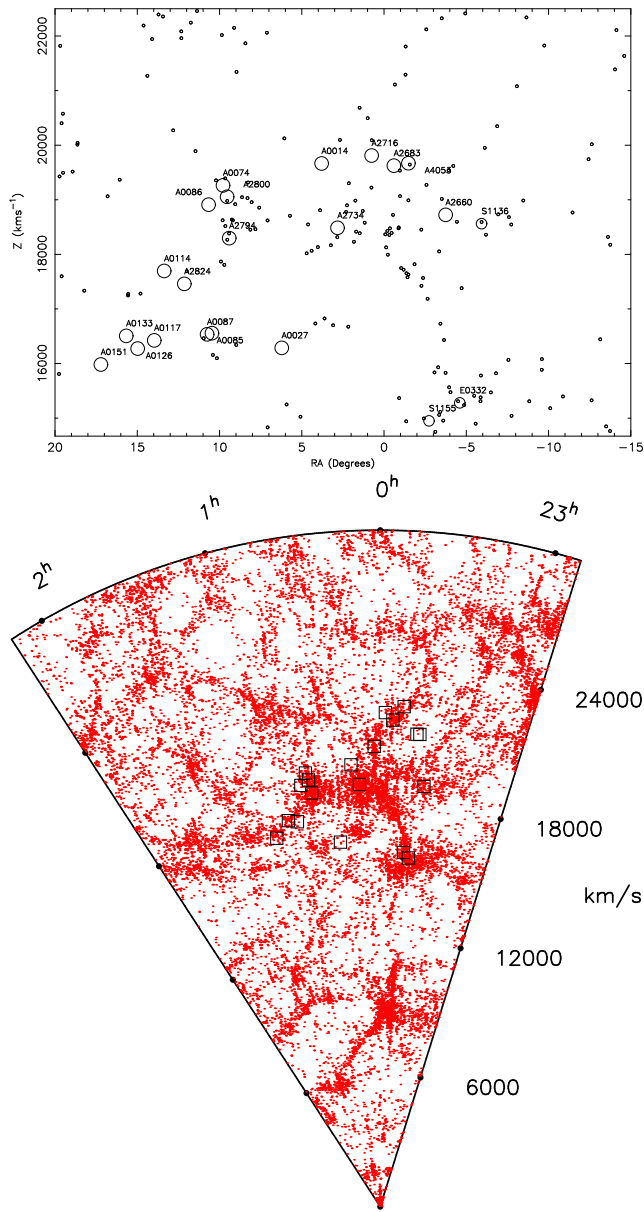

\psfig{figure=fig.2a.ps,angle=-90,width=\hsize}
\psfig{figure=fig.2b.ps,angle=-90,width=\hsize}
\medskip
 \caption{(a,top) Clusters of galaxies belonging to Pisces-Cetus
 supercluster, covering the declination range $-32^\circ\! <\!
 \delta<-9^\circ$.  The 2PIGG groups of galaxies are shown as
 little circles, but only in the declination range covered by the 2dFGRS
($-32^\circ\! <\! \delta<-25^\circ$). 
(b, bottom) Galaxies with redshifts (mostly from 2dFGRS, see \S3)
in the Pisces-Cetus supercluster region (Declination range
 $-32^\circ\! <\!  \delta<-20^\circ$).  The mean positions of the
 clusters in the supercluster are marked.  The 2dFGRS only goes up to
 $\delta<-25^\circ$, which is why some of the clusters don't seem to
 be associated with obvious concentrations.
\label{wedge}
}
\end{figure}

A plot of these clusters within the supercluster region and redshift
range, including the connecting MST, can be seen in
Fig.~\ref{fig:tree}a.  Also plotted are the groups of galaxies in the
region of the supercluster covered by the 2dFGRS [$\delta<-25^\circ$]
from the 2dFGRS Percolation-inferred Galaxy Group (2PIGG) catalogue 
\citep{eke2004}, which are seen to follow a
filamentary structure (though not necessarily the tree defined by the
MST), as expected from numerical simulations where groups are seen to
preferentially form on filaments and move along filaments to fall into
clusters.  For comparison, Fig.~\ref{fig:tree}b shows all galaxies
with spectroscopic redshifts from the SDSS and 2dFGRS catalogues,
where only galaxies within 1000 $km/s$ of the upper and lower redshift
bounds of the supercluster have been plotted. Overdensities of
galaxies are clearly visible in the regions around the richer clusters
and the filamentary structure of the supercluster is also visible.

Figs.~\ref{wedge} show the same galaxies, but using the redshift
information described below.  A combined look at Fig.~\ref{fig:tree}
and Fig.~\ref{wedge} reveals that the supercluster is made of two main
filaments. From Figs.~\ref{wedge}a, which plots redshift against Right
Ascension for the groups and clusters in the region, it is evident
that the filament containing A85 is at a lower redshift than the
longer filament, which has A133 at the near end and A2716 at the more
distant end. The poorer clusters E0332 and S1155 appear to be in a clump 
separated from both of the main
filaments, with a filament of groups joining these
three clusters to the principal filament.

%=section3==================================================

\section[]{Spectroscopic data}

Positions from the NASA/IPAC Extragalactic Database (NED) were taken as a
starting point for the cluster centres, and then they were located on
UK Schmidt Telescope (UKST) $B_J$ plates.  
The NED positions, particularly those from
\cite{Abell1989}, were largely inadequate.  Therefore, 
centres were determined from available X-ray data whenever possible
(see Appendix), and from a visual inspection of the appropriate UKST
plates otherwise. Our adopted coordinates can be found in
Table~\ref{tab-list}.

Redshifts were taken from all available sources including NED, ZCAT
(http://cfa-www.harvard.edu/$\sim$huchra/zcat/), the first data
release of the 6dF Galaxy Survey Database \citep{6df} [6dFGS DR1], the
2dF galaxy redshift survey
\citep{2dF-final} and the second release (DR2) of the Sloan Digital
Sky Survey \citep{sdss}.  The galaxies were initially extracted from
these catalogues within an Abell radius (2.14 $h_{70}^{-1}$ Mpc) of the
new cluster centres and filtered for repeated galaxies.

\subsection[]{Mean redshift and velocity dispersion}
\label{sec:red}

Mean recessional velocities for clusters that have redshifts available
for $\ge 4$ galaxies, within the Abell radius of 2.14 $h_{70}^{-1}$
Mpc, were calculated from the compilation described above, using an
iterative process clipping $3\sigma$ about the median, and finally
calculating the mean.  If $\ge 12$ galaxies with known redshift
remained, the velocity dispersion of the cluster and its error were
calculated, using a method that takes into account observational
errors on individual redshift measures \citep{danese1980}. The
resulting mean velocities, velocity dispersions and the number of cluster
members used to calculate them can be found in Table~\ref{tab-list}
for each cluster belonging to the Pisces-Cetus supercluster.

\begin{figure*}
\psfig{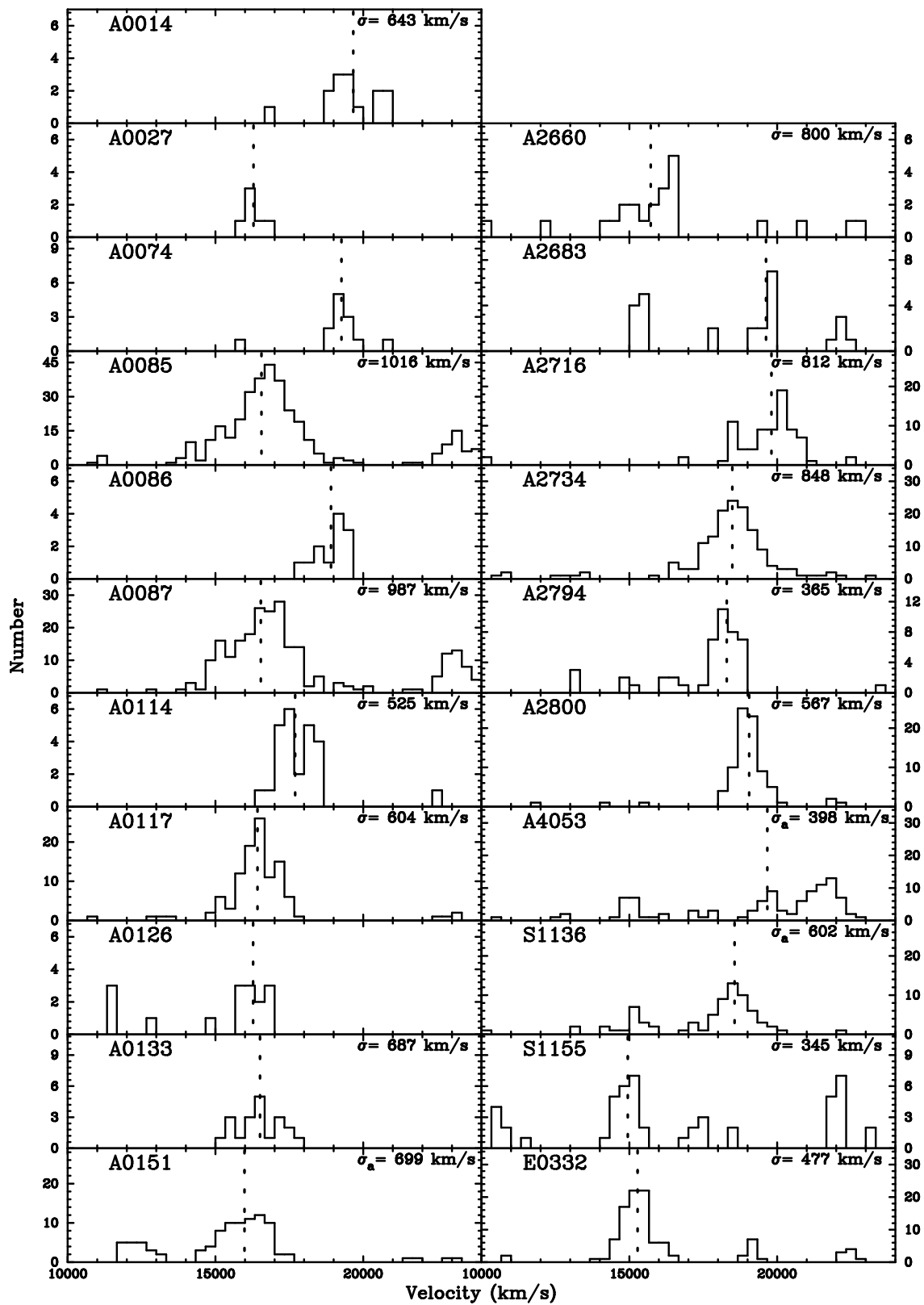}
\medskip
 \caption{ Radial velocities of galaxies lying within the Abell radius
 of 2.14 $h_{70}^{-1}$ Mpc of each galaxy cluster, belonging to the
 Pisces-Cetus supercluster, compiled from NED, 6dFGS, 2dFGRS and SDSS.
 For each cluster, the mean recessional velocity 
is shown as a dotted line, and
 for those with $\ge$12 redshifts, the velocity dispersion is shown on
 the upper right.  }
\label{hist}
\end{figure*}

Velocity histograms of each cluster can be seen in
Fig.~\ref{hist}. Where more than a peak is obviously present, the
velocity dispersion of the peak closest to the mean recessional
velocity of the supercluster has been adopted. However, in some cases,
membership remains ambiguous. For example, even though a large number
of galaxies were found with measured redshifts within the Abell radius
of A87, it is likely that a large fraction of them are in fact members
of A85, since the centres of these two are only $\sim$2 $h_{70}^{-1}$
Mpc apart.

Fig.~\ref{wedge}b shows all the galaxies with redshifts from 2dFGRS and
ZCAT in the declination range $-32^\circ\! <\! \delta<-20^\circ$.
The diagram shows clearly the filamentary nature of the supercluster
and the positions of the Abell clusters on the filaments.  There are
some clusters not obviously associated with overdensities-- these are
in the declination range not covered by the 2dFGRS.

%section=4=====================================

\section[]{The influence of the supercluster environment on
star formation in galaxies}

Studies of the environmental effects on galaxy evolution in the past
have shown that star formation is suppressed in the cores of rich
clusters \citep{dressler1980,couch1987,balogh1998}. More recently,
from the high-quality spectra of the 2dFGRS and SDSS archives, the
difference between the properties of star-forming and quiescent
galaxies have been further characterised, and shown to depend on the
local density of galaxies.
For instance, studies \citep{lewis2002,gomez2003,balogh04} identifying
star formation activity with the equivalent width of the $H_\alpha$
emission line $W\,(H_\alpha)$, show that the fraction of galaxies with
$W\,(H_\alpha)\!>\! 4$\AA\ steadily declines with increase in the
three-dimensional density of galaxies.
\citet{kauff04}
show similar effects, with the star formation history being
quantified from line indices measured elsewhere in the spectrum as
well, including the $H_\delta$ absorption line and the strength of the
$4000\AA$ break.

It is apparent from various
other studies \citep[\eg][]{miles04} that there is evidence for an
enhanced rate of galaxy mergers in the not-so-high-density
environments of groups, particularly in the poor ones, where the low
velocity dispersion of galaxies favours tidal interaction in a
dynamically sluggish system. This leads to, for instance, a depletion
in the intermediate-mass galaxies as evident from bimodal luminosity
functions. In such systems, we would expect the rate of merger-induced
star formation to be {\em enhanced} with respect to the field,
even though star formation is found to be quenched in high density
environments typical of rich clusters.

Indeed, in the hierarchically merging scenario supported by
large-volume numerical simulations such as \citet{virgo98}, much of
this merger-driven evolution in the history of
a galaxy is expected to
occur in small groups forming in and falling along a supercluster
filament towards regions of higher density. In contrast with the
behaviour of galaxies in the field and among the highest densities in
clusters, where one expects to find an abundance of quiescent
galaxies, we would expect to find an increase in the fraction of
galaxies with active star formation, induced by interaction, in the
groups that belong to a supercluster filament. Here we investigate this
effect by selecting a sample of galaxies that belong to groups
belonging to the part of the Pisces-Cetus supercluster in the 2dFGRS
volume, and comparing their properties with those of galaxies elsewhere.

\subsection[]{Rate of star formation: the $\eta$ parameter }

To investigate the variation in the rate of 
star formation rate with local environment, we used the 
spectral parameter $\eta$, which is found from a principal
components analysis \citep{madgwick2002} of the 2dFGRS spectra.
\citet{madgwick2003} found a tight correlation 
between $\eta$ and the equivalent width of the $H_{\alpha}$ emission
line, suggesting that $\eta$ is a measure of relative star formation.
Here we will look at galaxies with $\eta>0$, which corresponds to
$W\,(H_\alpha)\!\simgt \! 4$\AA, so that we can assume that these
galaxies have a significantly enhanced rate of star formation compared
to the mass of the older stellar population.

\subsection[]{Samples of galaxies in various environments}
\label{sec:samples}

Seven samples of galaxies
were complied from the 2dFGRS \citep{2dF-final} catalogue,
for which the  values of $\eta$ parameter were extracted.
These were:
\begin{enumerate}
\item{SC-PG}: Galaxies ($n=573$)
belonging to poor groups, extracted from the
2PIGG catalogue of groups found in the 2dFGRS \citep{eke2004},
which have $4\!\le\! N \!\le\! 10$ members, and lying within
the Pisces-Cetus supercluster region (Fig.~\ref{fig:tree});
\item{NSC-PG}: Same as above, but for poor 2PIGG groups that do not lie
in superclusters ($n=3822$).
This was achieved
by omitting all galaxies within a liberal radius of all supercluster
centres in the catalogue of \citet{somak05};
\item{SC-RG}: Galaxies ($n=6130$)
belonging to rich 2PIGG groups, 
which have $N\! >\! 10$ members, lying in the Pisces-Cetus
supercluster region;
\item{NSC-RG}: Same as above, but for groups not in superclusters 
(defined as in (ii));
\item{FIELD}: All galaxies ($n=44363$) 
in the 2PIGG catalogue not flagged as being a member of any group;
\item{SC-C}: Galaxies ($n=369$)
from Abell clusters that
are part of the Pisces-Cetus supercluster, falling within the
2dFGRS region.
\item{NSC-C}: Galaxies ($n=552$)
belonging to Abell clusters in the 2dFGRS regions,
with similar redshifts, but not belonging to any supercluster in the 
\citet{somak05} catalogue.

\end{enumerate}

\begin{figure*}
%\label{form1}
\centerline{
\psfig{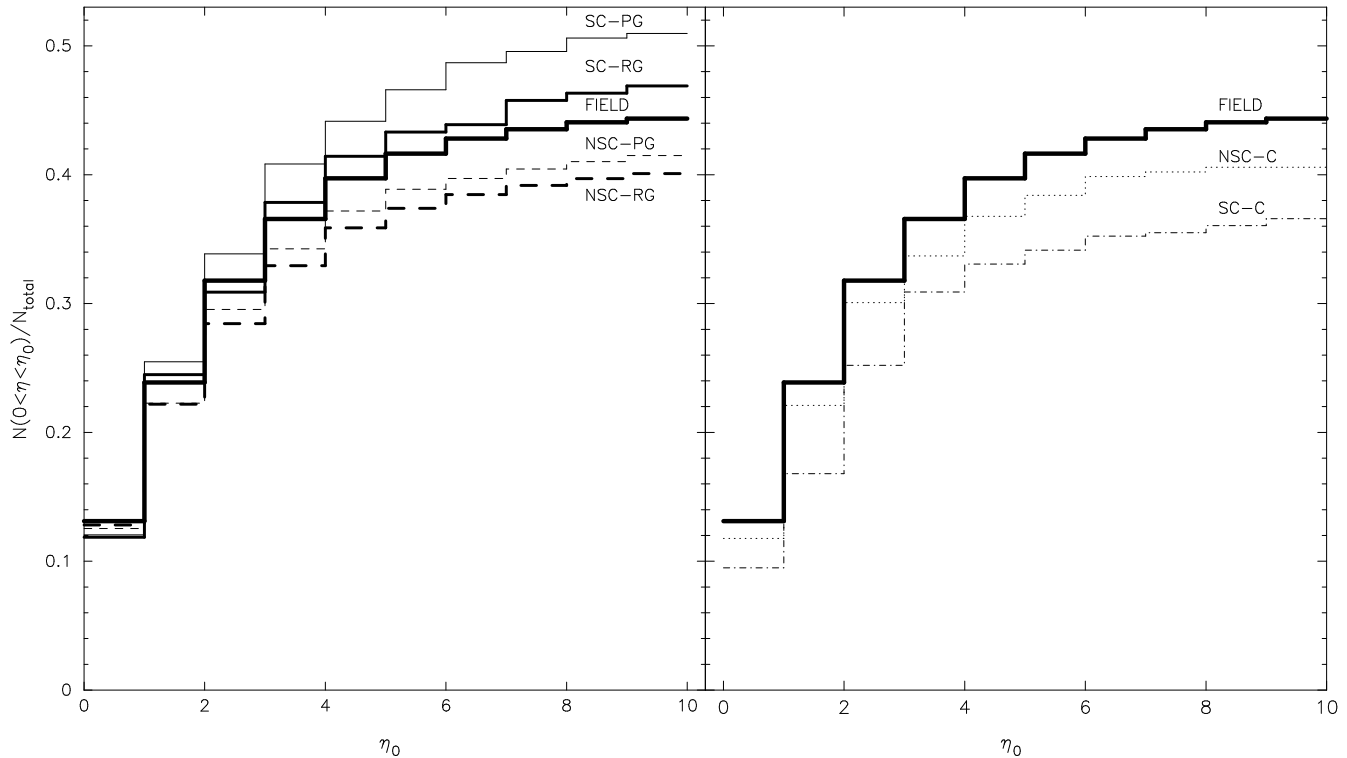}}
\medskip
 \caption{(Left, a) Cumulative distribution of the spectral parameter
 $\eta$ for galaxies with $0<\eta<\eta_0$, normalised by the total
 number of galaxies in each sample. A high value of the parameter
 $\eta$ indicates a higher width of the $H_{\alpha}$ emission line,
 and thus a higher rate of star formation. The thickest solid line
 refers to the FIELD sample (see section~\ref{sec:samples}). The
 thinnest and medium thick solid histograms refer to poor group
 galaxies within the supercluster, (SC-PG) and rich group galaxies
 within the supercluster (SC-RG) respctively.  The dashed histograms
 refer to poor group galaxies not within the supercluster (NSC-PG,
 thinner line), and to rich group galaxies not within the supercluster
 (NSC-RG, thicker line).  (Right, b) Cumulative distribution of
 $\eta$, as in (a), with the thick histogram being the field sample as
 in the left panel, the dash-dotted line representing cluster galaxies
 within the supercluster region (SC-C) and the dotted histogram
 standing for cluster galaxies not within the supercluster region
 (NSC-C). }
\label{eta1}
\end{figure*}

Since the rich ``groups'' and ``clusters'' are taken from different
lists, some ``cluster'' galaxies would be included within the ``rich
group'' samples, but this small number will be swamped by the much
larger number of group galaxies.

Assuming that $\eta>0$ 
for a galaxy signifies evidence of enhanced star formation, we
investigate the fraction of galaxies that have positive $\eta$ in the
various group and cluster samples, and compare them with the galaxies
in the field sample.  In Fig.~\ref{eta1}, we plot cumulative
histograms of the fraction of galaxies with a value of $\eta$ between
$0<\eta<\eta_0$ in each sample. The histograms are normalised by the
total number of galaxies in each sample.

To find a formal measure of the difference between the histograms, we
used both the familiar Kolmogorov-Smirnov (K-S) statistic and the
Kuiper statistic \citep{num-rec}. Since we are looking for differences
in our distributions in the high-$\eta$ tail, the Kuiper statistic
would in principle fare better than the K-S statistic in quantifying
differences.

It can be seen in Fig~\ref{eta1}a that there is a marked difference,
in the distribution of the $\eta$ parameter, 
between the group galaxies within and outside of the supercluster. For
both poor and rich groups, galaxies in those that belong to the
supercluster show evidence of enhanced star formation.  For the SC-PG
and FIELD samples, the K-S and Kuiper statistics indicate that the
probability of the samples to be drawn from the same parent
distribution is 0.01 and $10^{-8}$ respectively.  This would be
consistent with the filamentary structure of the supercluster leading
to an enhanced merger rate among galaxies belonging to the
supercluster filament.

For both SC and NSC groups, galaxies in poorer groups seem to have the
higher star formation rates.  This would support evidence of more
rapid merging within the poorer groups, which would have generally
lower velocity dispersions, which would in turn encourage tidal
interactions. This is in agreement with the difference found
between the luminosity functions of
galaxies by \citet{miles04} in X-ray bright and X-ray faint groups.

Even smaller probabilities are obtained when the SC-PG and SC-C
samples, \ie\ galaxies belonging to poor groups and to clusters in the
supercluster, are compared. On the other hand, the difference between
the SC-PG and SC-RG galaxies, however, is marginal, but
possibly significant, the
probability of being drawn from the same distribution is 0.53 and 0.1
according to the K-S and Kuiper statistics respectively.

It can be seen in Fig~\ref{eta1}b that the cluster galaxies have a
smaller percentage of their galaxies in the higher $\eta$ values than
the field sample, implying star formation is suppressed in the cluster
environment in agreement with
\citet{dressler1980,couch1987,balogh1998}.  However, the cluster
galaxies within the supercluster seem to have an even lower star
formation rate than the cluster galaxies outside of the
supercluster. This is the opposite trend to that found for the group
galaxies.

%=section5================================================================

\section[]{Radial Profiles of Light}

For all clusters belonging to the Pisces-Cetus supercluster, listed in
Table~1, all galaxies within a radius of $2.14\,h_{70}^{-1}$ Mpc
(equal to the Abell radius, defined as $1.5\,h_{100}^{-1}$ Mpc) of our
adopted centres were extracted from the Reliable Galaxies catalogue of
the SuperCosmos Science archive \citep{scos}.  Galaxies were radially
binned, with variable radius such that there was an equal number of
galaxies ($N\!\sim\! 10$) within each annulus. The luminosity per unit
area in each annulus was then calculated to plot the surface
brightness profiles, an example of which is shown in Fig.~\ref{fit}. A
10\% error is shown associated with each point.

\subsection[]{NFW Surface brightness profiles}

The radial surface brightness profiles were fit, using $\chi^2$
minimisation,
with the \citet{nfw:96} (NFW)
 model for luminosity density
\begin{equation}
\lambda(r)=
\frac{\delta_\lambda\,\rho_c}{\frac{r}{r_c}\left(1+\frac{r}{r_c}\right)^2},
\label{rho1} 
\end{equation} 
where $r_c=r_{200}/c$ is a characteristic radius,
$\rho_c=3H_0^2/8{\pi}G$ is the critical density ($H_0$ is the current
value of Hubble's constant), and $\delta_c$ and $c$ are two
dimensionless parameters, representing the mean overdensity and
central concentration of the radial profile respectively.  The mass
densities of the usual NFW profile are scaled to light density profiles
assuming light traces mass, at least in the visible cores of each
cluster, such that $\delta_\lambda=\delta_c/\Upsilon$, where
$\Upsilon\equiv M/L$ is the mean mass-to-light ratio of the cluster.

\begin{figure}
\psfig{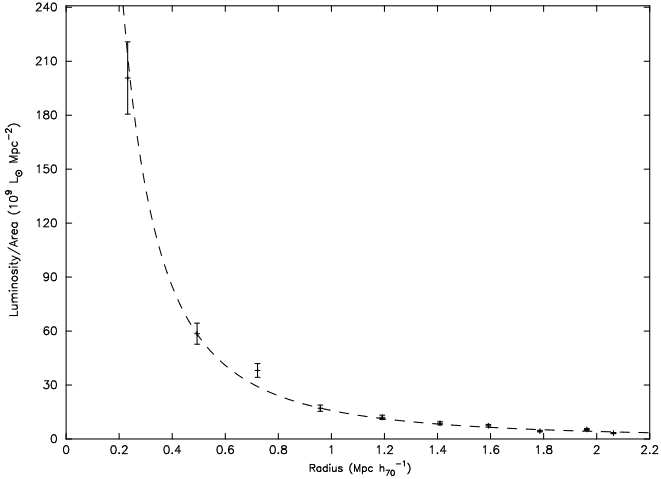}
\medskip
 \caption{The radial surface brightness profile for galaxies in the
 Abell 2734, with the best-fit projected NFW model
 (Eq.~\ref{eq:surfbrit}) superposed. Member galaxies are binned in
 annuli around the cluster centre, such that each annulus contains
an equal number of galaxies.
\label{fit}}
\end{figure}

This implies that the projected surface brightness
\begin{equation} 
\Lambda\,(R)=\frac{2\rho_c\delta_\lambda {r_c}}{x^2-1}f(x),
\label{eq:surfbrit} 
\end{equation}
with             
\begin{eqnarray}
f(x)=& 1-\frac{2}{\sqrt{x^2-1}}\arctan\sqrt\frac{x-1}{x+1} & \mbox{($x>1$)}\\ \nonumber
f(x)=& 1-\frac{2}{\sqrt{1-x^2}}\arctan\sqrt\frac{1-x}{1+x} & \mbox{($x<1$)}\\ \nonumber
f(x)=& 1 \hfil & \mbox{($x=1$)},                                               
\end{eqnarray}
where $x=r/r_c$, $R$ being the projected
radius from the centre of the cluster. 
This projected model is then fit to the observed
surface brightness profiles, with an additional constant term
to represent foreground/background galaxies. 
An example of such a fit can be found in Fig.~\ref{fit} for
the cluster Abell 2734.
The profiles are generally well fitted with 16 of the clusters having
reduced chi squared values of between 0.9 to 1.1, except for a
couple of cases with obviously deviant points.

\subsection[]{Virial mass estimates}

The mass 
within a radius $r$ from the centre of the cluster
can be derived from the modelled
velocity dispersion and spatial density profiles $\lambda\,(r)$ 
\citep{koranyi2000}.
On assumption of a constant mass-to-light ratio $\Upsilon\equiv M/L$,
the mass density $\rho\,(r)\equiv \Upsilon\, \lambda\,(r)$, following
(\ref{rho1}), results in the enclosed mass profile,
\begin{equation} 
M(<r)=4\pi\delta_c{\rho_c}r_c^3\big[\ln (1+\frac{r}{r_c})
   -\frac{r/r_c}{1+r/r_c}\big].
~\label{eq:mass2}
\end{equation}
The core radius $r_c$ has already been found in the fitting
of the surface brightness profile above. 
The quantity ${\delta}_c$ and $c$ are linked by the requirement that
the mean density within $r_{200}$ should be $200 {\times}
\rho_c$, \ie,
\begin{equation}
\delta_c=\frac{200}{3}\frac{c^3}{[ln(1+c)-c/(1+c)]},
~\label{delta1}
\end{equation} 
where $c=r_{200}/r_c$.

Assuming that the virial theorem holds, an estimate of $r_{200}$ can
be gained from 
\begin{equation} 
\frac{M_{200}r_v}{M_{\nu}r^3_{200}}=100H_0^2\frac{(1+z)^3}{3\sigma^2_r},
~\label{m200}
\end{equation} 
where $\sigma^2_r$ is the radial velocity dispersion of the cluster and $r_v$
is the approximate virial radius determined by
\begin{equation} 
~\label{vir1}
r_v\sim\frac{\pi}{2}\frac{N(N-1)}{\sum_i\sum_{i<j}R_{ij}^{-1}},
\end{equation}
where $N$ is the number of galaxies in the system.  The summation is
over all pairs of galaxies $(i,j)$, where the projected separation for
each pair is $R_{ij}$ \citep{Binney1987}.

The enclosed mass profile (\ref{eq:mass2}) can be used to find $M_{200}$
and $M_{\nu}$ in terms of ${\delta}_c$.  The calculated cluster masses
within the Abell radius of 2.14 $h_{70}^{-1}$ Mpc, and their core
radii and virial radii found from the model fitting, can be found in
Table~\ref{tab-list}.

The virial estimate will fail to represent the gravitating mass 
of a cluster which is very obviously in the process of rapid evolution,
\eg\ a collapsing or interacting system, since the virial
theorem does not strictly apply to systems whose moment of inertia is
significantly changing with time.  Abell~87 is obviously such a system
(see Appendix), so instead of measuring its virial mass, we have
assumed it to be equal in mass to Abell~85 in overdensity
calculations.

\subsection[]{A lower limit to the Mass of the Supercluster}

The sum of the individual virial masses of the clusters, as calculated
in the previous section, can be treated as a lower limit to the mass
of the supercluster. As described above, we assume that the mass of
Abell~87 is the same as that of Abell~85.  For the entire system
consisting of 19 Abell clusters and three non-Abell clusters, as
listed in Table~\ref{tab-list}, this amounts to 1.5$\times 10^{16}$
$h_{70}^{-1}$ M$_{\odot}$. This corresponds to the case where the
inter-cluster links in the MST are allowed to be $\le 25\,h^{-1}$
Mpc. However, as in this paper, if one considers inter-cluster links
of $20\,h^{-1}$ Mpc or less, the supercluster is made of at least two
distinct filaments.
\begin{enumerate}
\item For the filament containing A85, we use a cylindrical column of length
10.5~deg and cross-section of diameter 4~deg. For this volume of
6757~Mpc$^3$, we obtain the total mass of the clusters as 
$5.2 \times 10^{15}\,h_{70}^{-1}\,M_{\odot}$;

\item For the filament containing A133, using a cylindrical column of length
21~deg and cross-section of diameter 4~deg, amounting to
a volume of
19598~Mpc$^3$, we obtain the total mass of the clusters as 
$6.0 \times 10^{15}\,h_{70}^{-1}\, M_\odot$.
\end{enumerate}

With the critical density of the universe having a value of
$\rho_c=1.36\times 10^{11}\,h_{70}^{2}$ M$_{\odot}$ Mpc$^{-3}$, 
these values for the mass of the Supercluster 
represent matter overdensities of 
$\delta_M\equiv\Delta\Omega/\Omega_{crit}>$4.7 
(over $6757\,h_{70}^{-3}$~Mpc$^3$)
and $\delta_M>$1.3 (over $19598\,h_{70}^{-3}$~Mpc$^3$)
for the A85-related and A133-related
filaments respectively. These values are similar to those found
in the core of the Shapley supercluster, and the cosmological significance
of such overdensities in a supercluster filament has been discussed in
the literature \citep{shapnat,ettori97}. 
Given that these represent just the mass of the clusters
in the filament, and that the matter density of
the Universe is $\Omega_M\!\sim\! 0.3$, these amount to
very significant overdensities over very large volumes, larger
than is required for a given volume to turn around and collapse
upon itself.

\section{Conclusions}

We have used survey redshifts (mainly from 2dFGRS, 6dFGS, SDSS, NED
and ZCAT) and photometry (from SuperCosmos) of galaxies belonging to
the clusters belonging to the Pisces-Cetus supercluster to investigate
its nature, extent and orientation. The supercluster consists of two
main filaments, consisting of 11 and 5 clusters respectively, at mean
redshifts of $z=$0.0625 and 0.0545 respectively.  
Most of the former,
containing the cluster A133, lies in the 2dFGRS region, and gives us a
remarkable three-dimensional view of a string of clusters, with groups
delineating the filamentary structure.
The latter,
partially covered by the Sloan Digital sky survey, contains the
clusters A85 \& A87, for which independent evidence exists (from X-ray
observations) of collapse along the filament. 

For the galaxies lying in the region of the sky covered by the 2dFGRS,
we investigated the evidence of enhanced star formation in groups and
clusters within and outside the supercluster, using the $\eta$
parameter measured from principal components of the spectra, which has
been shown to correlate well with the equivalent width of the
$H_\alpha$ emission line. We found that while, as expected, cluster
galaxies have evidence of suppressed star formation compared to
galaxies in the field, a higher percentage of the galaxies in groups
belonging to the supercluster have enhanced star formation compared to
those in the field.  This would be consistent with the filamentary
structure of the supercluster leading to an enhanced merger rate,
which would in turn induce star formation.

Of these, from the $\eta$ values, it appears that galaxies in poor
groups in the supercluster have more enhanced star formation than
those in the richer groups.  This would imply that due to the enhanced
rate of galaxy mergers within the poorer groups, which in general have
lower velocity dispersions and thus the galaxies in them experience
more dynamical friction. This agrees with the difference of the
luminosity function of galaxies between X-ray bright and X-ray faint
groups \citep{miles04}.

Virial masses were calculated from NFW fits to surface brightness
profiles and velocity dispersions for the constituent clusters, the
sum of these masses being taken as a lower limit to the mass of the
supercluster.  These were found to be $5.2 \times
10^{15}\,h_{70}^{-1}\,M_{\odot}$ and $6.0 \times
10^{15}\,h_{70}^{-1}\,M_{\odot}$ respectively for the A85-filament and
A133-filament.  These correspond to mass overdensities of 
$\Omega/\Omega_{crit}\equiv
\delta_M>$4.7 and $\delta_M>$1.3, over volumes of almost
$6750\,h_{70}^{-3}$~Mpc$^3$ and $19600\,h_{70}^{-3}$~Mpc$^3$)
respectively, for the A85-related and A133-related filaments.
These
values indicate that in both these branches, it is likely that a large
fraction of the supercluster has already turned around and collapsed
upon itself.

We started this discussion with the debate concerning the scale of the
largest bound structures in the Universe, and the scale of homogeneity
of the Universe. Ever since the discovery of the ``Great Wall''
\citep{geller1989}, 
which seemed to be a structure almost $<200\,h_{70}^{-1}$ Mpc long,
the discovery of comparable structures have been claimed
\citep[e.g.,][]{bagchi02,brand03,miller04}.  To be cosmologically significant,
it is necessary to show that these large structures are
gravitationally bound, which is difficult to demonstrate.  In this
work, the two supercluster filaments we consider are respectively
$100\,h_{70}^{-1}$ Mpc and $50\,h_{70}^{-1}$ Mpc long. A large
fraction of these filaments seem to be bound entities, showing
evidence of ongoing evolution, and interaction-induced star formation.
This is consistent with statistical analyses of the LCRS and 2dF
redshift surveys, which indicate that the scale of homogeneity sets in
beyond the scale of $\sim 100\,h_{70}^{-1}$ Mpc or so.

\section*{Acknowledgements}

We would like to thank Suketu Bhavsar for the development of the
minimum spanning tree software used to find the supercluster
members. We would also like to thank all the people involved in
creating the 2dFGRS, 6dFGS, SDSS and SuperCosmos surveys, which form
the source of our data. Our gratitude goes to an anonymous referee who
provided helpful comments.  Finally, we would like to thank the
creators of NASA/IPAC Extragalactic Database (NED), the first port of
call in many cases.  NED is operated by the Jet Propulsion Laboratory,
California Institute of Technology, under contract with the National
Aeronautics and Space Administration.

\begin{table*}
\begin{center}
\caption{Clusters of galaxies belonging to the Pisces-Cetus Supercluster
\label{tab-list}
}
\footnotesize
\begin{tabular}{lcccrclccccccc} 
\hline
Cluster  &R.A.  &Dec&cz    &\# z   &$L_{\rm x}$      &T      &Rich &Survey &$\sigma_r$    &$r_{200}$  &$r_v$  &$M_{\rm NFW}$\\ 
&(J2000)&(J2000)&km~s$^{-1}$&&$\times 10^{44}$ erg/s&keV&&&
km~s$^{-1}$&Mpc&Mpc&$10^{15}\,M_\odot$\\
\hline
A0014   &00 15 10.9 &-23 52 56 &19665            &13  &   0.51\rlap{$_b$}             &2.8\rlap{$_{b,*}$}  &0    &-      &$643^{+185}_{-100}$  &2.33  &6.23   &1.1\\
A0027   &00 24 52.1 &-20 43 14 &16287            & 6  &   \llap{$<$}0.028\rlap{$_d$}  &1.0\rlap{$_*$}      &0    &-      &                     &1.46  &6.53   &\\
A0074   &00 39 08.8 &-22 19 49 &19264            &11  &   0.021\rlap{$_a$}            &1.1\rlap{$_*$}      &0    &-      &                     &1.62  &6.16   &\\
A0085   &00 41 50.8 &-09 18 18 &16556            &294 &   12.47\rlap{$_c$}            &6.2\rlap{$_a$}       &1    &S      &$1016^{+44}_{-39} $  &3.12  &5.57   &1.8\\
A0086   &00 42 40.0 &-21 48 01 &18911            &12  &   0.094\rlap{$_a$}            &1.3\rlap{$_*$}      &0    &-      &                     &2.00  &6.27   &\\
A0087   &00 43 01.0 &-09 47 36 &16535            &194 &                               &                     &1    &S      &$987^{+54}_{-46}$    &      &       &  \\
A0114   &00 53 27.5 &-21 44 42 &17696            &24  &   0.055\rlap{$_a$}            &1.4\rlap{$_*$}      &0    &-      &$525^{+100}_{-65}$   &2.10  &6.29   &0.6\\
A0117   &00 55 51.9 &-09 59 09 &16424            &101 &   0.26\rlap{$_a$}             &1.5\rlap{$_*$}      &0    &S      &$604^{+47}_{-38}$    &2.23  &6.23   &0.7\\
A0126   &00 59 53.8 &-14 14 41 &16274            &11  &   0.027\rlap{$_d$}            &1.2\rlap{$_*$}      &1    &-      &                     &1.85  &6.13   &\\
A0133   &01 02 41.2 &-21 52 45 &16507            &20  &   3.81\rlap{$_c$}             &3.8\rlap{$_a$}       &0    &-      &$687^{+146}_{-90}$   &2.57  &6.08   &1.2\\
A0151   &01 08 51.1 &-15 24 23 &15982            &73  &   0.70\rlap{$_b$}             &3.1\rlap{$_{b,*}$}   &1    &-      &$699^{+66}_{-52}$    &2.35  &5.55   &0.9\\
A2660   &23 45 16.0 &-25 50 09 &18724            &17  &   0.049\rlap{$_d$}            &1.7\rlap{$_*$}      &0    &0.73   &$800^{+189}_{-111}$  &2.54  &6.35   &1.6\\
A2683   &23 57 36.7 &-25 36 44 &19625            &11  &                               &                     &0    &0.88   &                     &0.93  &6.71   &\\
A2716   &00 03 01.3 &-27 08 11 &19810            &77  &                               &                     &0    &0.85   &$812^{+75}_{-59}$    &2.66  &6.20   &1.0\\
A2734   &00 11 19.5 &-28 51 41 &18486            &141 &   2.22\rlap{$_c$}             &4.4\rlap{$_{b,*}$}  &1    &0.82   &$848^{+55}_{-46}$    &2.84  &6.07   &1.6\\
A2794   &00 37 38.1 &-30 48 59 &18298            &34  &                               &                     &1    &0.97   &$365^{+56}_{-40}$    &1.69  &6.02   &0.4\\
A2800   &00 38 09.4 &-25 04 51 &19054            &79  &                               &                     &1    &0.81   &$567^{+52}_{-41}$    &2.08  &6.31   &0.5\\
A2824   &00 48 34.0 &-21 20 36 &17460\rlap{$_e$} & 7  &   0.47\rlap{$_b$}             &2.8$_{b,*}$         &0    &-      &                     &      &       &    \\
A4053   &23 54 46.0 &-27 40 18 &19670            &26  &                               &                     &1    &0.88   &$398^{+72}_{-48}$    &1.12  &5.99   &\\
S1136   &23 36 16.9 &-31 36 20 &18562            &54  &   1.25\rlap{$_b$}             &3.7$_{b,*}$         &0    &0.89   &$602^{+68}_{-51}$    &2.41  &6.39   &1.0\\
S1155   &23 49 44.0 &-29 07 13 &14950            &33  &                               &                     &0    &1.00   &$345^{+57}_{-42}$    &1.51  &6.78   &0.3\\
ED332   &23 41 35.5 &-29 14 11 &15280            &84  &                               &                     &     &1.00   &$477^{+43}_{-35}$    &2.24  &6.36   &0.6\\                                                  	 
\hline
\end{tabular}
\end{center}
\noindent
{\footnotesize
Columns: 
(1) Name of cluster (A*** and S*** \citet{Abell1989}, 
ED332 \citet{Lumsden1997}; 
(2,3) Equatorial J2000 centres (we use centroids of
 X-ray emission where available, see Appendix); 
(4) Redshift (see \S\ref{sec:red}); 
(5) Number of
cluster member redshifts used to find the mean cluster redshift 
((e) \citet{struble1999} and \S\ref{sec:red}); 
(6) Bolometric X-ray luminosities ((a)\citet{Jones1999}; 
(b) \citet{Cruddace2002}; (c) \citet{reiprich2002}; (d) \citet{ledlow2003}); 
(7) Temperature ($\ast$=estimated from $L_X-T$ relation \citet{Osmond2004}); 
(8) Abell richness (\citet{Abell1989}); 
(9) Survey coverage [S=Sloan, Number= completeness in 2dFGRS field];
(10) Velocity dispersion (see \S\ref{sec:red}); 
(11) $r_{200}$ in Mpc (see Eq.~(\ref{m200})); 
(12) Virial radius (see Eq.~(7)); 
(13) Cluster virial mass within an Abell radius 
               (2.14 $h_{70}^{-1}$ Mpc) (see Eq.~(\ref{eq:mass2})). 
}
\end{table*}

\appendix
\section[]{X-Ray observations}

\begin{figure*}
\centerline{
%14 14 151 213
\psfig{figure=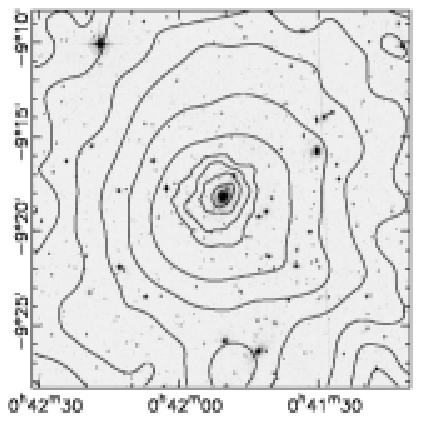,width=0.33\hsize,angle=0,bbllx=20bp,%
  bblly=50bp,bburx=151bp,%
  bbury=170bp,clip=}
\psfig{figure=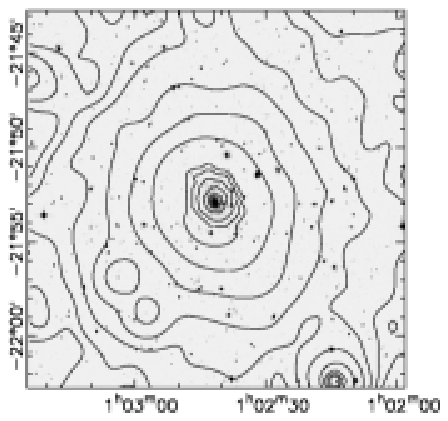,width=0.33\hsize,angle=0,bbllx=20bp,%
  bblly=50bp,bburx=151bp,%
  bbury=170bp,clip=}
\psfig{figure=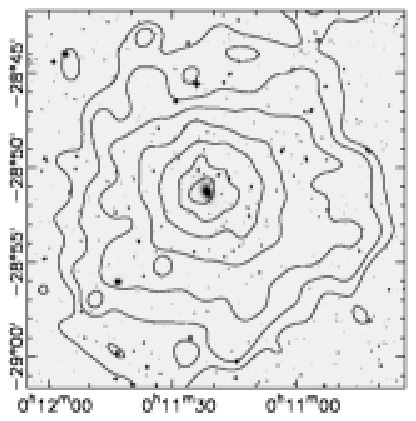,width=0.33\hsize,angle=0,bbllx=20bp,%
  bblly=50bp,bburx=151bp,%
  bbury=170bp,clip=}}
%14 14 315 223
\centerline{\psfig{figure=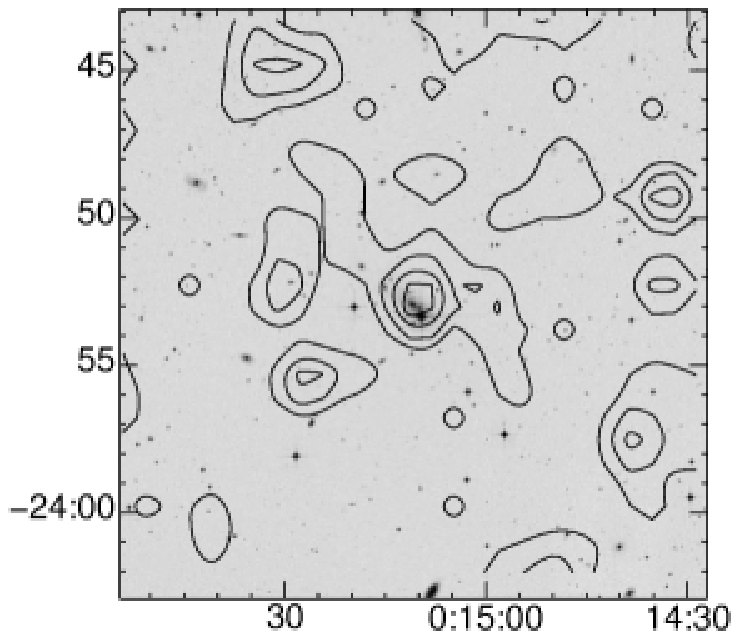,width=0.25\hsize,angle=0,%
  bbllx=35bp,bblly=15bp,bburx=255bp,%
  bbury=210bp,clip=}
%14 14 256 177
\psfig{figure=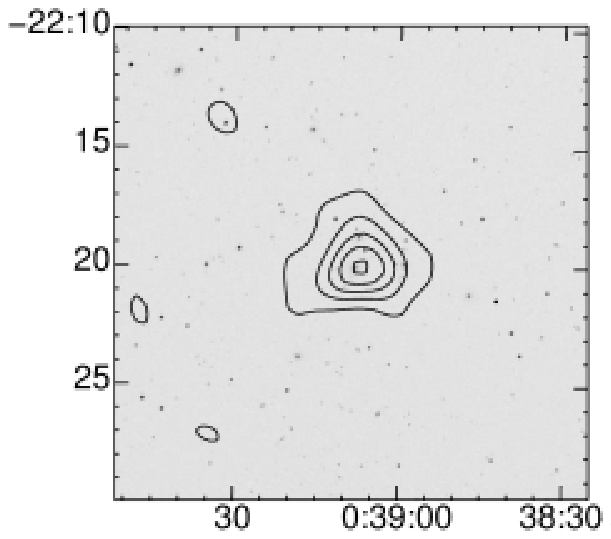,width=0.25\hsize,angle=0,bbllx=15bp,%
  bblly=15bp,bburx=195bp,%
  bbury=177bp,clip=}
%14 14 315 222
\psfig{figure=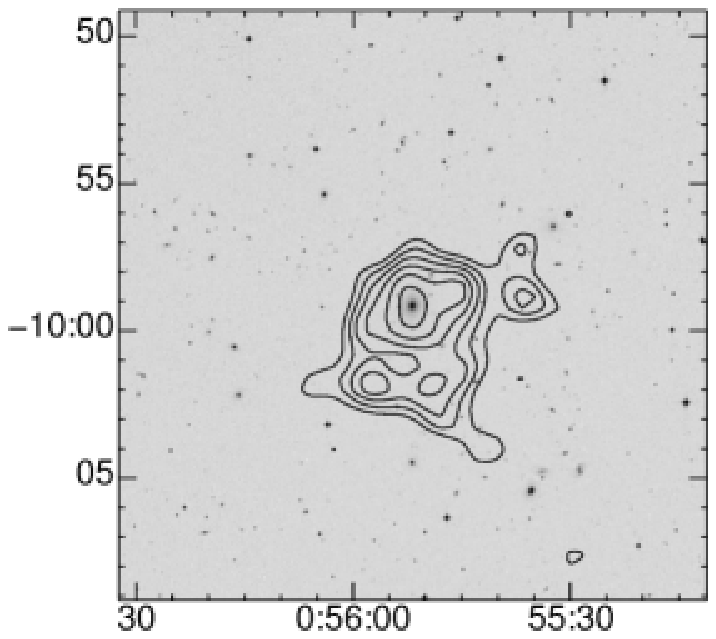,width=0.25\hsize,angle=0,bbllx=40bp,%
  bblly=15bp,bburx=255bp,%
  bbury=210bp,clip=}
%14 14 315 223
\psfig{figure=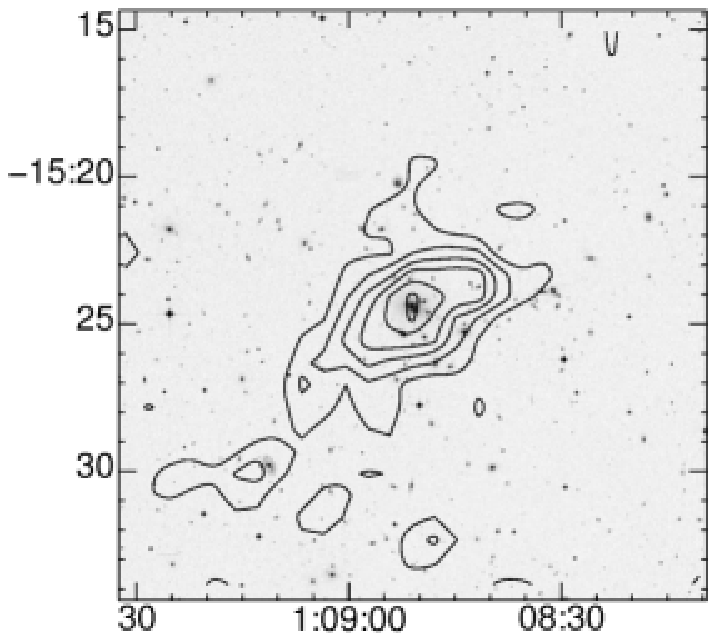,width=0.25\hsize,angle=0,bbllx=40bp,%
  bblly=15bp,bburx=255bp,%
  bbury=215bp,clip=}}
\centerline{%
%14 14 315 222
\psfig{figure=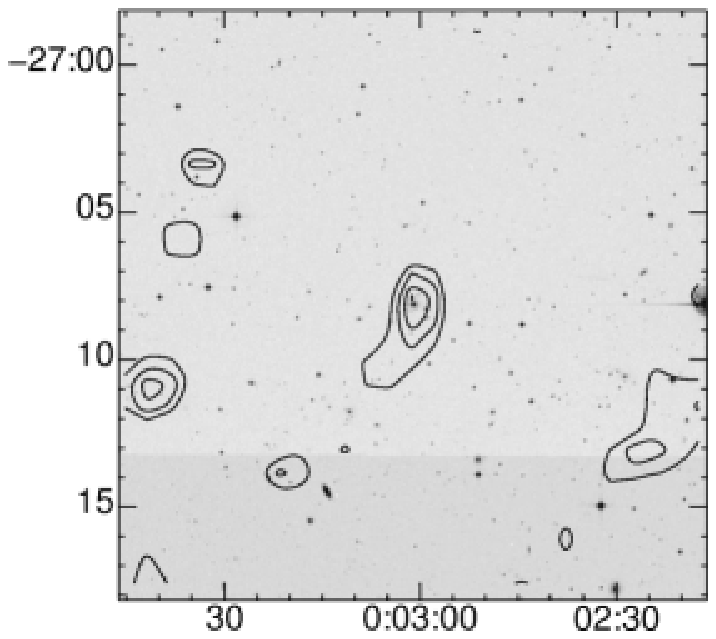,width=0.25\hsize,angle=0, 
 bbllx=40bp,bblly=15bp,bburx=260bp,%
  bbury=210bp,clip=}
\psfig{figure=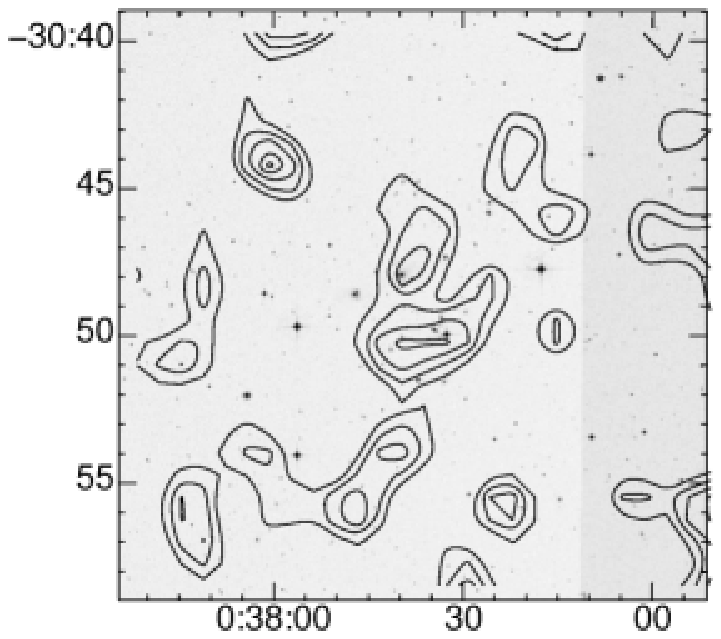,width=0.25\hsize,angle=0,
 bbllx=40bp,bblly=15bp,bburx=260bp,%
  bbury=210bp,clip=}
\psfig{figure=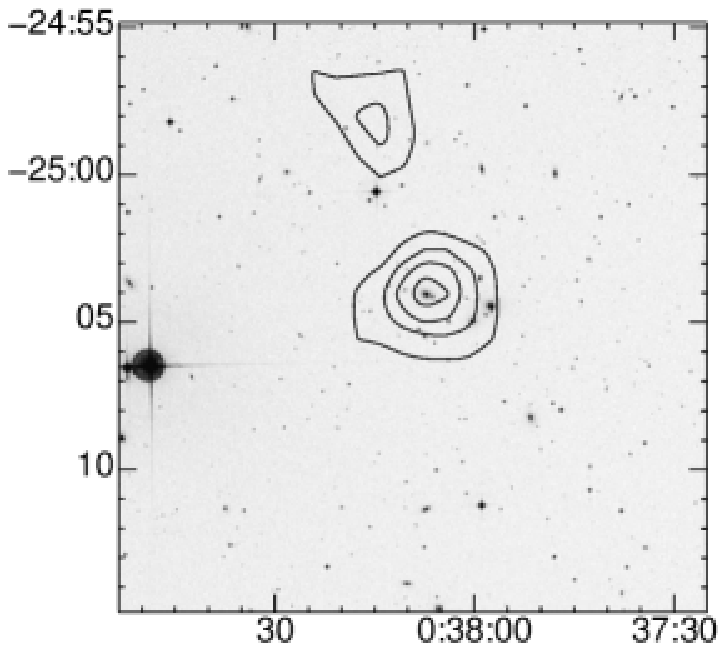,width=0.25\hsize,angle=0,
 bbllx=40bp,bblly=15bp,bburx=260bp,%
  bbury=210bp,clip=}
\psfig{figure=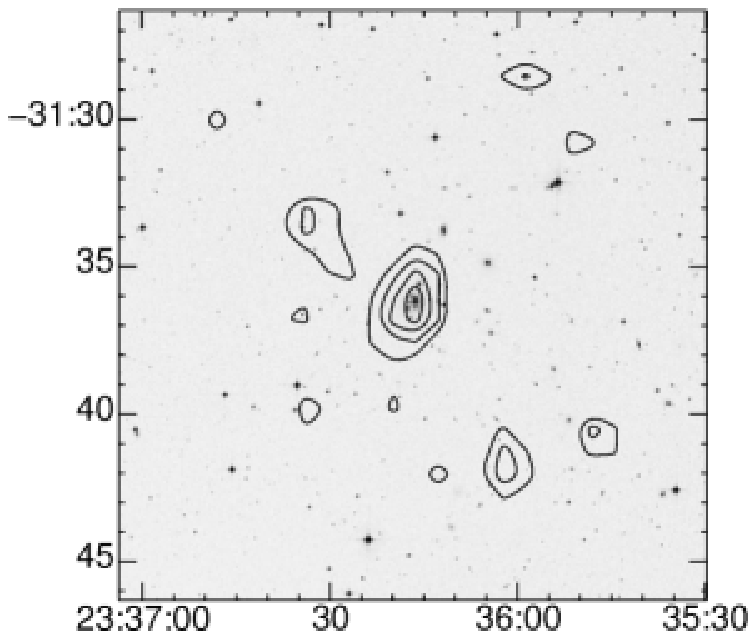,width=0.25\hsize,angle=0,
bbllx=40bp,bblly=15bp,bburx=260bp,%
  bbury=210bp,clip=}}
\medskip
 \caption{
(Top row) X-ray contours from pointed ROSAT PSPC observations, superposed 
on optical DSS images (blue) of the clusters
Abell~85, 
Abell~133 and 
Abell~2734. (Middle row) Similar X-ray contours of
Abell~14 (RASS),
Abell~74 \&
Abell~117 (Einstein IPC),
Abell~151 (RASS), and (Bottom row)
of
Abell~2716,
Abell~2794,
Abell~2800, and 
Abell~S1136 (all RASS), superposed on optical DSS images.
}
\label{xcon1}
\end{figure*}

We reduced the available archival ROSAT PSPC pointed observations of
A85, A133 and A2734, as well as those of the others that were detected in
the ROSAT all-sky survey (RASS) and the EINSTEIN observations, using
the ASTERIX software package (http://www.sr.bham.ac.uk/asterix-docs).
The smoothed contours of X-ray flux in the 0.1-2.4 keV range can be
seen as overlays on optical (blue) DSS images in Fig.~ (\ref{xcon1}).
X-ray luminosities from literature were converted into bolometric luminosities,
with the software package PIMMS, using a Raymond-Smith model.
The temperature of the plasma, when not found in the literature, was
estimated from the velocity dispersion-temperature relation of
\citet{Osmond2004} and a suitable H~I column density. The luminosity
and temperature values are listed in Table~\ref{tab-list}.

\section[]{Notes on individual clusters}

\paragraph*{Abell~14}
Our adopted position from a visual inspection of DSS images is offset by
$\Delta$R.A.$=-3^{\prime\prime}$ and $\Delta$Dec$=+22^{\prime\prime}$ from the 
original Abell position. \citet{Abell1989} noted that the
brightest galaxy and the 3rd brightest are spirals.
\citet{baier2003} analysed a sample of 974 COSMOS/UKST southern sky object 
catalogue galaxies from the cluster and found substructures in the cluster.

\paragraph*{Abell~27, 86, 114, 117, 126}  
Our adopted positions from a visual inspection of DSS images are offset by
(A~27): $\Delta$R.A.$=+3^{\prime\prime}$ and $\Delta$Dec$=-50^{\prime\prime}$;
(A~86): $\Delta$R.A.$=+11^{\prime\prime}$ and $\Delta$Dec$=-18^{\prime\prime}$;
(A~114): $\Delta$R.A.$=-11^{\prime\prime}$ and $\Delta$Dec$=-4^{\prime}$;
(A~117): $\Delta$R.A.$=-9^{\prime\prime}$ and $\Delta$Dec$=+2^{\prime} 
39^{\prime\prime}$;
(A~126): $\Delta$R.A.$=+7^{\prime\prime}$ and $\Delta$Dec$=-1^{\prime} 
53^{\prime\prime}$ from 
the original Abell position. 
    
\paragraph*{Abell~74}
Our adopted position from RASS X-ray emission and confirmed by a 
visual inspection of DSS images is offset by
$\Delta$R.A.$=+16^{\prime\prime}$ and $\Delta$Dec$=-1^{\prime} 
19^{\prime\prime}$ from the original Abell position.

\paragraph*{Abell~85, Abell~87}  
Our adopted position for Abell~85
from a visual inspection of DSS images is offset by
$\Delta$R.A.$=+13^{\prime\prime}$ and $\Delta$Dec$=+2^{\prime} 
18^{\prime\prime}$ from the original Abell position. 
\citet{Giovannini2000} note that  
radio emission is detected from the central cD galaxy of A85, and two
head-tail radio galaxies are also visible. 
%They also show that two
%extended regions of radio emission with no obvious optical
%identification are present. 
XMM-Newton observations \citep{durret2003} 
have shown that A85 is a strong X-ray source
with an extended 4 Mpc filament to the southeast while A87 is
not detected. \citet{kempner2002} 
investigate the south subcluster of A85 to the north of this filament,
and confirm that this subcluster is merging with the main cluster, 
showing evidence of significant disruption, possibly associated
with non-thermal inverse Compton emission \citep{bagchi98}/

\paragraph*{Abell~133}
Our adopted position is taken from \citet{Jones1999}, and is offset by
$\Delta$R.A.$=+3^{\prime\prime}$ and $\Delta$Dec$=-4^{\prime}
51^{\prime\prime}$ from the original Abell position.
\citet{edge1991} note the presence of a 
bright cD galaxy with multiple nuclei at the centre of a cooling flow
X-Ray cluster.
\citet{Slee2001} point out the
cluster has a radio relic, which is a diffuse steep-spectrum source
with no optical identification. 
\citet{fujita2002} found a tongue of X-ray emission 
extending the the northwest of the cD galaxy with no optical
counterpart. The X-ray spectrum showed the emission to be thermal and
at a lower temperature than the ambient gas.  \citet{fujita2004} find
evidence of a weak merger shock, and suggest it to be the origin of
the tongue-like X-ray structure.

\paragraph*{Abell~151}
Our adopted position from a visual inspection of DSS images is offset by
$\Delta$R.A.$=-1^{\prime\prime}$ and $\Delta$Dec$=+37^{\prime\prime}$ 
from the original Abell position. 
\citet{proust1992} studied the kinematics of the cluster, and found
that it is composed of three subclusters at $z=$0.054,
0.041 and 0.11. They suggest that the background one at z=0.11 is likely
to be part of a supercluster including A86, A131, A159, A166 and A183. They 
also find a southern substructure which they conclude is a foreground group.
\citet{girardi1998} show that
the cluster consists of two separate clumps, A151a, 44 galaxies
($\sigma_r=714$ km/s) and A151b, 7 galaxies $\sigma_r=385$ km/s).

\paragraph*{Abell~2660, Abell~2683, Abell~2716, Abell S1136} 
Our adopted position from a visual inspection of DSS images is offset
by (A2660): $\Delta$R.A.$=-2^{\prime\prime}$ and
$\Delta$Dec$=+8^{\prime} 9^{\prime\prime}$ and (A2683):
$\Delta$R.A.$=+3^{\prime\prime}$ and $\Delta$Dec$=-3^{\prime}
26^{\prime\prime}$ and
(A2716): $\Delta$R.A.$=+10^{\prime\prime}$ and $\Delta$Dec$=+2^{\prime} 
7^{\prime\prime}$, 
(AS1136):
$\Delta$R.A.$=+3^{\prime\prime}$ and $\Delta$Dec$=+4^{\prime\prime}$ 
from the original Abell position.

\paragraph*{Abell~2734}
Our adopted position taken from \citet{Cruddace2002} 
and confirmed by a visual inspection of DSS images is offset by
$\Delta$Dec$=+37^{\prime\prime}$ from the original Abell position. 
   
\paragraph*{Abell 2794, Abell S1155}
Our adopted position taken from \citet{Lumsden1992} and confirmed by a
visual inspection of DSS images is offset by (A2794):
$\Delta$R.A.$=+1^{\prime}5^{\prime\prime}$ and
$\Delta$Dec$=+12^{\prime} 31^{\prime\prime}$ and (AS1155):
$\Delta$R.A.$=-27^{\prime\prime}$ and $\Delta$Dec$=-5^{\prime}
5^{\prime\prime}$ from the original Abell position.

\paragraph*{Abell~2800} 
Our adopted position from RASS X-ray emission and confirmed by a visual 
inspection of DSS images is offset by
$\Delta$R.A.$=+11^{\prime\prime}$ and $\Delta$Dec$=39^{\prime\prime}$ from 
the original Abell position. 

\paragraph*{Abell 4053}
\citet{Mazure1996} list a foreground group: 
$z=0.0501$ from 7 redshifts, $\sigma_r=731$ km/s)

\label{lastpage}

\clearpage

%\bsp
\end{document}